\documentclass[prd,nofootinbib,preprint,superscriptaddress]{revtex4}

\usepackage{graphicx} 
\usepackage{bm} 
\usepackage{epsfig}
\usepackage{amsmath}

\newcommand{\be}{\begin{equation}}
\newcommand{\ee}{\end{equation}}
\newcommand{\sigv}{\langle \sigma v \rangle}

\newcommand{\capdef}{}
\newcommand{\mycaption}[2][\capdef]{\renewcommand{\capdef}{#2}%
       \caption[#1]{{\footnotesize #2}}}
\makeatletter
\renewcommand{\fnum@table}{\textbf{\tablename~\thetable}}
\renewcommand{\fnum@figure}{\textbf{\figurename~\thefigure}}
\makeatother

\begin{document}

\title{Phenomenology of  Dark Matter annihilation into a
  long-lived intermediate state\vspace*{1cm}}

\author{\bf Ira Z.\ Rothstein}
\email{izr_AT_andrew.cmu.edu}
\affiliation{Carnegie Mellon University, Dept.\ of Physics, Pittsburgh PA  15213, USA}

\author{\bf Thomas Schwetz}
\email{schwetz_AT_mpi-hd.mpg.de}
\affiliation{Max-Planck-Institute for Nuclear Physics, PO Box 103980, 69029 Heidelberg, Germany}

\author{\bf Jure Zupan\vspace*{3mm}}
\email{jure.zupan_AT_cern.ch}
\affiliation{Theory Division, Physics Department, CERN, CH-1211 Geneva 23,
Switzerland}
\affiliation{Faculty of mathematics and physics, University of
  Ljubljana, Jadranska 19, 1000 Ljubljana, Slovenia\vspace*{3mm}}

\begin{abstract}
  \vspace*{5mm} We propose a scenario where Dark Matter (DM)
  annihilates into an intermediate state which travels a distance
  $\lambda \equiv v/\Gamma$ on the order of galactic scales and then
  decays to Standard Model (SM) particles. The long lifetime disperses
  the production zone of the SM particles away from the galactic
  center and hence, relaxes constraints from gamma ray observations on
  canonical annihilation scenarios. We utilize this set up to explain
  the electron and positron excesses observed recently by PAMELA,
  ATIC and FERMI. While an explanation in terms of usual DM annihilations seems
  to conflict with gamma ray observations, we show that within the
  proposed scenario, the PAMELA/ATIC/FERMI results are consistent with the
  gamma ray data. The distinction from decay scenarios is discsussed
  and we comment on the prospects for DM production at LHC. The
  typical decay length $\lambda \gtrsim 10$~kpc of the intermediate
  state can have its origin from a dimension six operator suppressed
  by a scale $\Lambda \sim 10^{13}$~GeV, which is roughly the seesaw
  scale for neutrino masses.
  \end{abstract}

\vskip .2in
\maketitle

\section{Introduction}

An exciting and very plausible possibility is that dark matter (DM)
interacts non-gravitationally. A very active
ongoing program exists to search for such interactions both directly and
indirectly. Indirect evidence could arise from cosmic ray signals
originating from areas of high dark matter density --- the galactic
halos and cores.  Decaying or annihilating DM can act as an
additional source of cosmic ray fluxes on Earth, and
could be seen as an excess above the expected flux.  The  channel
in which the excess could show up depends on how DM couples to standard model
(SM) particles.

DM decays and annihilations into charged particles inevitably lead to
enhanced gamma ray fluxes due to internal and final state
bremsstrahlung. In both cases the gamma ray spectrum will be
correlated with the energy spectrum of charged particle, and will be
the same (up to overall energy scale) for the same final
states. However, the angular distribution of gamma rays in the sky
will differ between the two cases.  Since the annihilation rate scales
with $\rho^2$ ($\rho$ being the DM density), the gamma ray signal will
be more highly peaked toward the galactic center due to the rise in
the density profile for small $r$. The decay rate, on the other hand,
scales linearly with $\rho$ and therefore the gamma ray signal is
less peaked.

In this paper we consider an alternative scenario, where the DM annihilates
into long lived particles (LLP). This setup will lead to a gamma ray
signature which can interpolate between the signatures of DM decays and
annihilation.  The prolonged decay lifetime of LLP effectively smears out
the distribution of gamma rays.  To be concrete we will consider this
scenario in the context of DM explanation of the positron excess found by
PAMELA~\cite{Adriani:2008zr}, ATIC~\cite{atic} and FERMI~\cite{Abdo:2009zk}.
The PAMELA experiment which sees a rise in the positron-to-electron
ratio in cosmic rays for energies between 10 and 100~GeV corroborates and
extends the range of earlier data~\cite{olddata}, while ATIC sees a rise and
then a striking fall near 800~GeV in the power-law for the combined electron
and positron flux. The rise at lower energies, below the bump, corroborates
previous data~\cite{olddata2} while the bump at 800~GeV had never been seen
before. Recent data from the FERMI satellite do not confirm this bump,
while an excess of the electron and positron flux over the expectations from
common cosmic ray models is still there. FERMI data indicate a smooth spectrum
falling approximately as $E^{-3}$ till about 800~GeV. Above this energy data from
the HESS telescope~\cite{Collaboration:2008aaa, Aharonian:2009ah} indicate a
much steeper slope of the electron/positron flux. Although it seems
quite reasonable to believe that this data has an explanation in terms of
nearby young pulsars \cite{taylor, backer, Hooper:2008kg, Yuksel:2008rf,
Profumo:2008ms, Shaviv:2009bu, Malyshev:2009tw, Hall:2008qu} (or some of the
data could be spurious), it will serve as a useful tool in highlighting the
features of the LLP scenario. In particular, the data exemplify a case where
there is a need for an increased annihilation cross section $ \langle \sigma
v \rangle \sim 10^{-22}\,\rm cm^3 /s$ (relative to the typical cross section
$\langle \sigma v \rangle \sim 3 \times 10^{-26} \, \rm cm^3 /s$ needed for
$\Omega_\mathrm{DM}\sim 0.2$), while there is no evidence of excess in the
gamma ray flux in the direction of the galactic center.  Standard
annihilation scenarios with such large cross sections seem to
conflict~\cite{Bertone:2008xr, Bell:2008vx,Bergstrom:2008ag, Meade:2009rb}
with the results from the HESS telescope~\cite{Aharonian:2006wh,
Aharonian:2006au} for commonly assumed DM profiles.  We will show that, by
allowing the annihilation products to be long lived, the constraints from
gamma rays can be avoided.

To be specific, let us consider the situation where a DM particle
$\chi$ annihilates into an intermediate state $\phi$ which
subsequently decays into standard model particles
\be\label{eq:react}
\chi\chi \to \phi\phi \to \rm 2\,SM\,2\,\overline{SM} \,.
\ee
As an example we will assume in the following that $\phi$
decays into muons, 
\be\label{eq:react2}
\phi \to \mu^+\mu^- \,,
\ee
though our arguments do not rely on the specific decay mode of $\phi$. We
concentrate on this mode since it is the relevant one for the PAMELA/ATIC
data (while not changing our main conclusions other annihilation modes may
be preferred when including new FERMI data \cite{Meade:2009iu}). In
contrast to \cite{Pospelov:2008jd, ArkaniHamed:2008qn, Cholis:2008qq,
Nomura:2008ru, Cholis:2008wq, Fox:2008kb, Chen:2009dm} we now assume that
$\phi$ is a long-lived particle (LLP), such that it will propagate over
galactic distances.\footnote{Phenomenologically, annihilations into rapidly
decaying particles are challenging to distinguish from pure annihilation
\cite{Meade:2009rb,Mardon:2009rc} and thus we will not consider them
distinct in this paper.} For the moment we will not concern ourselves with
whether or not the DM particle $\chi$ is a thermal relic. We will return to
this issue at the end of the paper in section~\ref{sec:relic}.  In
Sec.~\ref{sec:gamma} we present a general discussion of the LLP scenario and
discuss the effects on the production of Standard Model particles from DM
annihilations focusing on gamma rays. In Sec.~\ref{sec:data} we apply this
idea to the recent electron and positron flux data from PAMELA, ATIC,
FERMI and HESS and show that they can be made consistent with gamma ray
observations from HESS even for rather cuspy DM profiles. In
Sec.~\ref{sec:conclusion} we summarize our results, discuss possible
distinguishing features of the scenario, mention prospects to see DM at LHC,
and speculate on a possible connection between the high scale physics
responsible for the decay of $\phi$ and the suppression of neutrino masses
via the seesaw mechanism.

\section{Gamma rays from DM annihilations via a long-lived state}
\label{sec:gamma}

\subsection{General discussion}

Let a $\phi$ be produced at $\vec{r} = 0$. Then the probability that
it decays in a volume element $d^3r$ around $\vec{r}$ is
\be\label{eq:def-lambda}
\frac{1}{4\pi \lambda} \, 
\frac{e^{- r / \lambda }}{r^2} \, d^3r\,,
\quad\text{with}\quad
\lambda \equiv
\frac{\gamma\beta}{\Gamma}\, 
\ee
where $\Gamma$ is the decay rate of $\phi$, $\lambda$ the
corresponding decay length in the laboratory frame, while the velocity
of $\phi$ in the laboratory frame, $\beta=v/c$, and the relativistic
gamma factor are
\be\label{eq:gamma-beta}
\gamma = \frac{E_\phi}{m_\phi} = \frac{m_\chi}{m_\phi} 
\,,\quad
\beta = \frac{p_\phi}{E_\phi} = \sqrt{1 - \frac{1}{\gamma^2}} =
\sqrt{1 - \frac{m_\phi^2}{m_\chi^2}} \,,
\ee 
where $m_{\chi,\phi}$ are the masses of $\chi$ and $\phi$. In the last
equalities above we used the fact that $\chi$ is non-relativistic.

Let us now calculate the photon flux resulting from final state
radiation of the process shown in Eqs.~\ref{eq:react} and
\ref{eq:react2}. The differential photon flux $\Phi_\gamma$ in a solid
angle $\Delta\Omega$ is given by 
\be\label{eq:spectrum}
\frac{d\Phi_\gamma}{dE_\gamma} = 
2 \, \langle \sigma v\rangle
\int_{\Delta\Omega} d\Omega 
\int_\mathrm{l.o.s.} ds 
\int d^3r' \frac{1}{2} \frac{\rho^2(r')}{m_\chi^2}
\frac{1}{4\pi \lambda} 
\frac{e^{- |\vec{r} - \vec{r'}| / \lambda }}
{|\vec{r} - \vec{r'}|^2} \,
\frac{1}{2\pi}
\frac{d^2 N_\gamma}{dE_\gamma d\cos\theta} \,.
\ee
Here $\rho(r)$ is the energy density of DM particles $\chi$, $\sigv$
is the averaged annihilation cross section of $\chi$, and we have
assumed that $\chi$ is a Majorana particle, a choice we shall adopt
throughout this work.  The factor of $1/2$ accounts for the fact that
the incoming particles are identical, as it has not been included in
the definition of $\langle \sigma v \rangle$.  $d^2N_\gamma/(dE_\gamma
d\cos\theta)$ is the double differential gamma spectrum per $\phi$
decay. It can be obtained by boosting the isotropic spectrum in the
$\phi$ rest frame (RF), where
\be\label{eq:sp-NR}
\left.\frac{d^2 N_\gamma}{dE_\gamma d\cos\theta} \right|_\mathrm{RF} 
= 
\left.\frac{1}{2} \frac{d N_\gamma}{dE_\gamma} \right|_\mathrm{RF} \,.
\ee
The factor 2 in Eq.~\ref{eq:spectrum} accounts for the fact that we
will have two $\phi$ decays per annihilation.

In Eq.~\ref{eq:spectrum} we use coordinates with the origin at the
galactic center, and $\chi\chi$ annihilate at the point $\vec{r'}$ and
$\phi$ decays at $\vec{r}$. Then $\theta$ is the angle between the
line of sight (l.o.s.) and the direction of motion of $\phi$:
\be
\cos\theta = -\frac{\hat{s} \cdot (\vec{r} - \vec{r'}) }{ |\vec{r} -
  \vec{r'}|}
\ee
where $\hat{s}$ is the unit vector along the line of sight. 
In the following we are going to compare the predicted photon flux to
data from the HESS telescope, using observations of the galactic
center (GC)~\cite{Aharonian:2006wh} within a cone of solid angle
$10^{-5}$. In this case we have
\be\label{eq:GC}
\Delta\Omega_\mathrm{GC} = 2\pi (1 - \cos\Delta\psi) = 10^{-5} 
\,,\quad
\int_{\Delta\Omega} d\Omega = \int_0^{2\pi} d\varphi 
\int_0^{\Delta\psi} \sin\psi d\psi \,,
\ee
\be
r = \sqrt{r_\odot^2 + s^2 - 2r_\odot s \cos\psi} \,,\quad
\hat{s} = (\cos\psi, \cos\varphi\sin\psi, \sin\varphi\sin\psi) \,,
\ee
where $r_\odot \simeq 8.5$~kpc is the distance of the solar system
from the galactic center. Furthermore, we use the HESS observation of
the galactic ridge (GR)~\cite{Aharonian:2006au}. This is a rectangular
region of the sky with galactic latitude and longitude of $|l| \le \Delta l
= 0.8^\circ$, $|b| \le \Delta b = 0.3^\circ$, where the inner region
corresponding to the GC defined above has been subtracted.  For the GR
we have
\be\label{eq:GR}
\Delta\Omega_\mathrm{GR} \approx 
4 \Delta l \Delta b - \Delta\Omega_\mathrm{GC}
\approx 2.8\times 10^{-4}
\,,\quad
\int_{\Delta\Omega} d\Omega \approx
4 \int_0^{\Delta l} dl \int_0^{\Delta b} db \,,
\ee
\be\label{eq:GR2} 
r = \sqrt{r_\odot^2 + s^2 - 2r_\odot s \cos l \cos b} \,,\quad \hat{s}
\approx \left(1 - \frac{l^2 + b^2}{2}, l, b\right) \,. 
\ee 
To compare with HESS observations \cite{Aharonian:2006au}, one additionally
needs to subtract from the photon flux observed in the GR region $|l| \le
0.8^\circ$, $|b| \le 0.3^\circ$ the average photon flux in the region
$0.8^\circ < |b| < 1.5^\circ$, a procedure that was used for the background
subtraction in~\cite{Aharonian:2006au}.

\bigskip 

Before we specialize to the case of non-relativistic $\phi$ and
proceed with our discussion we mention briefly some limiting cases of
Eq.~\ref{eq:spectrum}. First, we consider the limit $\Gamma \to \infty$
(or $\lambda\to 0$). In this case the integrand in Eq.~\ref{eq:spectrum}
will be non-zero only for $\vec{r} = \vec{r'}$. Hence we can replace
$\rho(r') \to \rho(r)$, which can be pulled out of the $d^3r'$
integral. Using for the remaining $d^3r'$ integral
\be
\int d\Omega' \, \frac{1}{2\pi}
\frac{d^2 N_\gamma}{dE_\gamma d\cos\theta} =
\frac{d N_\gamma}{dE_\gamma}  \,,
\ee
one obtains
\be
\left. \frac{d\Phi_\gamma}{dE_\gamma} \right|_{\Gamma\to\infty}
= 
\frac{\langle \sigma v\rangle}{4\pi} \frac{d N_\gamma}{d E_\gamma}
\int_{\Delta\Omega} d\Omega 
\int_\mathrm{l.o.s.} ds \,
\frac{\rho^2(r)}{m_\chi^2} \,.
\ee
As expected we recover the standard expression for the bremsstrahlung
photon flux from annihilation $\chi\chi \to \ell^+\ell^-+n\gamma$,
apart from a factor 2, since in our case there are twice as many
photons because of two $\phi$ decays per $\chi$ annihilation (see
Eq.~\ref{eq:react2}).

Second, we consider the case of highly relativistic $\phi$. Then one
has $\gamma \gg 1$ and $\beta \approx 1$, while the photons are
collimated with $\phi$ within an opening angle of size $1/\gamma$.  This
implies that there is only a contribution to the integral when
$(\vec{r}-\vec{r'})$ is aligned with the line of sight, and the
$d^3r'$ integration reduces to another integration along the line of
sight:
\begin{eqnarray}
\left.\frac{d\Phi_\gamma}{dE_\gamma}\right|_\mathrm{relat.} &=& 
\frac{\langle \sigma v\rangle}{4\pi}
\frac{d N_\gamma}{dE_\gamma }
\int_{\Delta\Omega} d\Omega 
\int_0^\infty ds'  
\int_0^{s'} ds \,
\frac{\rho^2(r')}{m_\chi^2} \,
\frac{e^{- (s'-s) / \lambda}}{\lambda} \nonumber\\
&=& 
\frac{\langle \sigma v\rangle}{4\pi}
\frac{d N_\gamma}{dE_\gamma}
\int_{\Delta\Omega} d\Omega 
\int_\mathrm{l.o.s.} ds' \,
\frac{\rho^2(r')}{m_\chi^2}
\left(1 - e^{- s'/ \lambda}\right) \label{eq:relat}
\end{eqnarray}

As a reference, we also give the expression for the photon flux from
decaying dark matter (see, e.g.~\cite{Nardi:2008ix}) 
\be\label{decay}
\left. \frac{d\Phi_\gamma}{dE_\gamma} \right|_{\rm decay}
= 
\frac{\Gamma}{4\pi} \frac{d N_\gamma}{d E_\gamma}
\int_{\Delta\Omega} d\Omega 
\int_\mathrm{l.o.s.} ds \,
\frac{\rho(r)}{m_\chi} \,,
\ee
where $\Gamma$ is the DM decay width. The important difference between
decaying and annihilating DM is that the DM density enters linearly in
the first case, and quadratically in the second, which has important
observational implications discussed below.

\subsection{The effective DM profile for non-relativistic intermediate state $\phi$}

In order to simplify the calculations we specialize from now on to the
case of non-relativistic $\phi$, as the generic effects will not
change for the case of a relativistic $\phi$. The effect of an
intermediate LLP with a suitable decay length $\lambda$ will always be
to flatten out the distribution of SM particle production with respect
to the DM distribution, independently of its $\gamma$.  This follows
from the structure of Eq.~\ref{eq:relat} or the general expression
Eq.~\ref{eq:spectrum}. The general case would deserve a dedicated
quantitative study which is beyond the scope of the present work.

For non-relativistic $\phi$ one has $\gamma \approx 1$, $\beta \ll 1$,
and the decay of $\phi$ is isotropic. In particular, the photon
spectrum is given by Eq.~\ref{eq:sp-NR}. The spectrum can be pulled
out of the integration and the effect is a smearing of the density
distribution on the scale $\lambda$. One may define an effective
density distribution:
\begin{eqnarray}
\rho^2_\mathrm{eff}(r) &=&
\int d^3r' \rho^2(r')
\frac{1}{4\pi \lambda} 
\frac{e^{- |\vec{r} - \vec{r'}| / \lambda }}
{|\vec{r} - \vec{r'}|^2} \nonumber\\
&=&
\frac{1}{2\lambda r} \int_0^\infty dr' r'
\rho^2(r')
\left[ 
  \mathrm{Ei} 
    \left(-\frac{r + r'}{\lambda}
    \right) -
  \mathrm{Ei} 
    \left(-\frac{|r - r'|}{\lambda}
    \right)
\right] \,, \label{eq:rho_eff}
\end{eqnarray}
with the exponential integral
\be
\mathrm{Ei}(x) \equiv -{\rm PV} 
\left( \int_{-x}^\infty dt \frac{e^{-t}}{t} \right)\,,
\ee
where PV denotes principal value. Using the effective DM profile in
Eq.~\ref{eq:rho_eff} we can write the photon flux from a given solid angle
$\Delta\Omega$ as
\be\label{eq:F_gamma}
\frac{d\Phi_\gamma}{dE_\gamma} = 
\frac{\sigv}{4\pi} \,\frac{r_\odot \rho_\odot^2}{m_\chi^2}\,
\frac{dN_\gamma}{dE_\gamma}\,  J \, \Delta\Omega \,
\ee
where the dimensionless $J$-factor is defined by
\be\label{eq:J}
J = \frac{1}{\Delta\Omega} \int_{\Delta\Omega} d\Omega 
\int_\mathrm{l.o.s.}
\frac{ds}{r_\odot}\,\frac{\rho_\mathrm{eff}^2(r)}{\rho_\odot^2} \,.
\ee
For the numerical calculations in this work we use $\rho_\odot =
0.3\,\rm{GeV\,cm^{-3}}$ and $r_\odot = 8.5$~kpc, and for the DM
density $\rho(r)$ we always assume a NFW
profile~\cite{Navarro:1995iw} 
\be
\rho(r) = \rho_\odot \, \frac{r_\odot}{r} 
\left( \frac{1 + r_\odot/r_s}{1 + r/r_s} \right)^2 \,,
\ee
with $r_s = 20$~kpc. Eq.~\ref{eq:F_gamma} has the same form
as the photon flux from standard annihilations
$\chi\chi\to\mu^+\mu^-+n\gamma$, apart from a factor 2 from the decay of
two $\phi$'s, as mentioned above.

In the non-relativistic case the two parameters $m_\phi$ and $\Gamma$
appear only in the particular combination
$\lambda=\gamma\beta/\Gamma$, see Eqs.~\ref{eq:def-lambda} and
\ref{eq:gamma-beta}. Therefore, apart from the two parameters $\sigv$
and $m_\chi$ of the standard annihilation scenario, we have now
effectively one additional parameter corresponding to the decay length
of $\phi$ in the rest frame of the galaxy. Numerically one has
\be\label{eq:param}
\lambda = \frac{\gamma\beta}{\Gamma} \approx \frac{\sqrt{2\delta}}{\Gamma} 
\approx 1.4\,{\rm kpc} \, \left(\frac{\delta}{0.01}\right)^{1/2}
\left(\frac{\tau}{10^{12}\,\rm s}\right) \,,
\ee
with $\delta \equiv (m_\chi - m_\phi)/m_\chi \ll 1$ and $\tau =
1/\Gamma$.

\begin{figure}
\centering 
\includegraphics[width=0.65\textwidth]{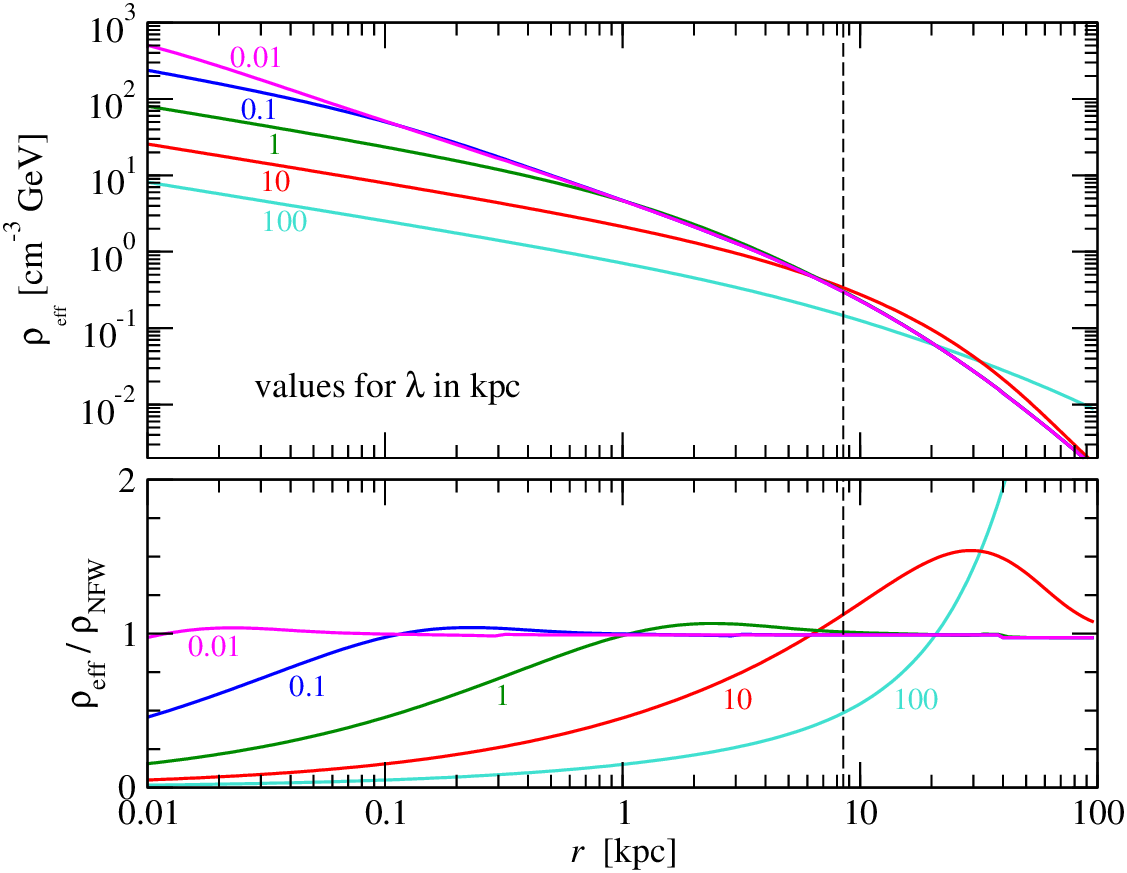} 
  \mycaption{\label{fig:rho_eff} Upper panel: effective DM density profiles as
    defined in Eq.~\ref{eq:rho_eff} for various values of
    $\lambda$. Lower panel, effective DM profiles
    relative to the NFW profile (corresponding to $\lambda = 0$).}
\end{figure}

In the LLP scenario the source of SM particles from DM annihilations is
proportional to $\rho_\mathrm{eff}^2(r)$, while in the case of
standard annihilations it is proportional to $\rho^2(r)$. This means
that via the propagation of the intermediate state $\phi$ over
galactic distances we decouple to some extent the production of SM
particles and associated gammas from the DM distribution. 
Fig.~\ref{fig:rho_eff} shows the effective DM profile for various
values of $\lambda$. From the figure we find that for $r \lesssim
\lambda$ we suppress $\rho_\mathrm{eff}(r)$ with respect to $\rho(r)$
since the $\phi$ had no time to decay yet, while for $r\gtrsim\lambda$
we obtain a $\rho_\mathrm{eff}$ slightly larger than the original DM
profile.  The relative size of this over-production increases with
$\lambda$. Finally, at large distances $r \gg \lambda$ all $\phi$'s
have decayed and we recover the NFW profile.

\begin{figure}
\centering 
\includegraphics[width=0.55\textwidth]{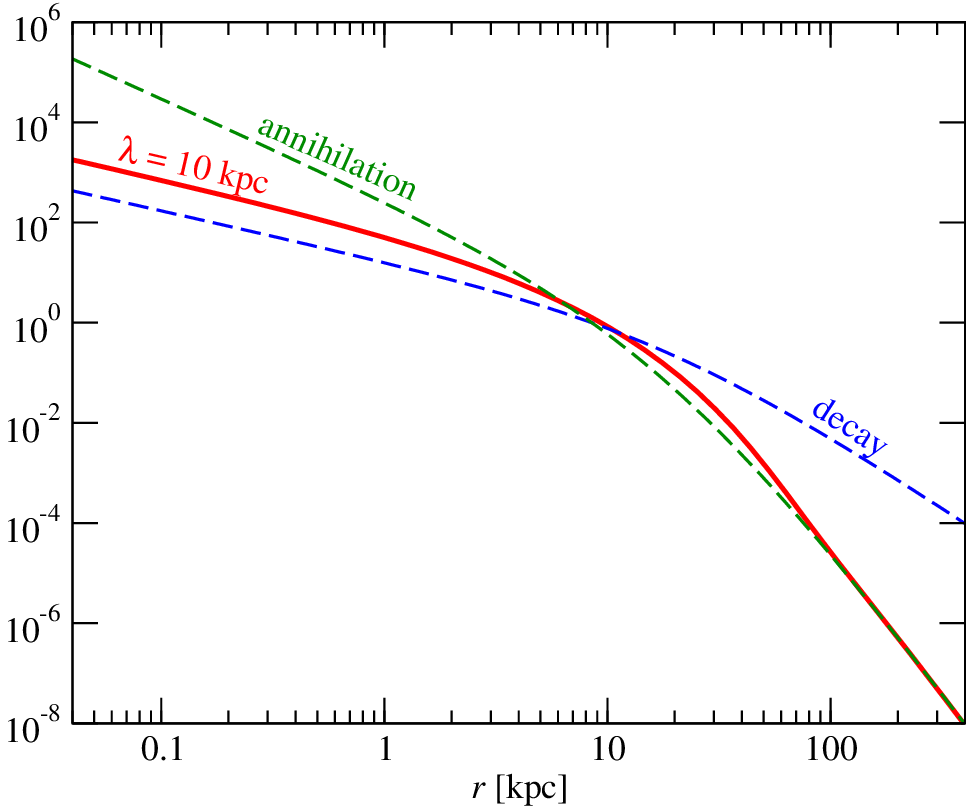}
\mycaption{\label{fig:slope} Interpolation between the SM particle
  source terms for DM annihilation and decay. The dashed curves show
  $\rho^2_\mathrm{NFW}(r)/\rho_\odot^2$ for annihilation and
  $\rho_\mathrm{NFW}(r)/\rho_\odot$ for decay, whereas the solid curve
  corresponds to $\rho_\mathrm{eff}^2(r)/\rho_\odot^2$ for DM
  annihilation into a long lived intermediate state with a decay
  length of $\lambda = 10$~kpc.}
\end{figure}

One can understand this behavior qualitatively from the definition of
$\rho_\mathrm{eff}$ in Eq.~\ref{eq:rho_eff}. From the first line of
this equation it follows that for $r \gg \lambda$ the exponential is
non-zero only for $\vec{r} \approx \vec{r'}$ and one obtains
$\rho_\mathrm{eff}(r) \to \rho(r)$. On the other hand, for $r \ll
\lambda$ and for profiles $\rho(r) \propto r^{-\gamma}$ ($\gamma
\gtrsim 1$) one has
roughly
\be
\left. \rho^2_\mathrm{eff}(r)\right|_{r\ll \lambda} \propto 
\int  \frac{d^3r'}{{r'}^{2\gamma} \, |\vec{r} - \vec{r'}|^2} =
\int  \frac{d\Omega \, dr'}
{{r'}^{2\gamma-2} \, |\vec{r} - \vec{r'}|^2} \propto \frac{1}{r^{2\gamma - 1}} \,,
\ee
where the last relation follows just from dimensional analysis. Hence,
the slope of the photon source term at $r\ll\lambda$ is reduced by one
power with respect to $\rho^2(r)$. Since for a NFW profile $\gamma =
1$, we find for $r\ll\lambda$ that $\rho_\mathrm{eff}^2(r) \propto
\rho_\mathrm{NFW}(r) \propto 1/r$. Note that for DM decay the source
of gamma rays is proportional to $\rho(r)$, in contrast to the
$\rho^2(r)$ for annihilations. Therefore, for a NFW profile our
scenario exactly interpolates between DM decay at $r\ll\lambda$ and
annihilation at $r\gg\lambda$. This behavior is shown in
Fig.~\ref{fig:slope}. Note that the parameter $\lambda$ controls the
absolute value of $\rho_\mathrm{eff}^2$ at small $r$. For $\lambda
\simeq 40$~kpc one would match exactly the profile for DM decay in the
particular (somewhat arbitrary) normalization used in
Fig.~\ref{fig:slope}. However, for such large $\lambda$ the asymptotic
$\rho^2(r)$ behavior at large $r$ is reached only at distances much
larger than the size of our galaxy. We emphasize that the {\it slope}
at small $r$ is independent of $\lambda$, and hence it is a generic
prediction of our scenario that for profiles with $\gamma \simeq 1$ at
$r \lesssim r_\odot$ the gamma ray signal looks like DM decay from the
inner region of the galaxy (cf.\ Eq.~\ref{decay}), whereas it mimics
DM annihilation when looking away from the galactic center.

\begin{figure}
\centering 
\includegraphics[width=0.6\textwidth]{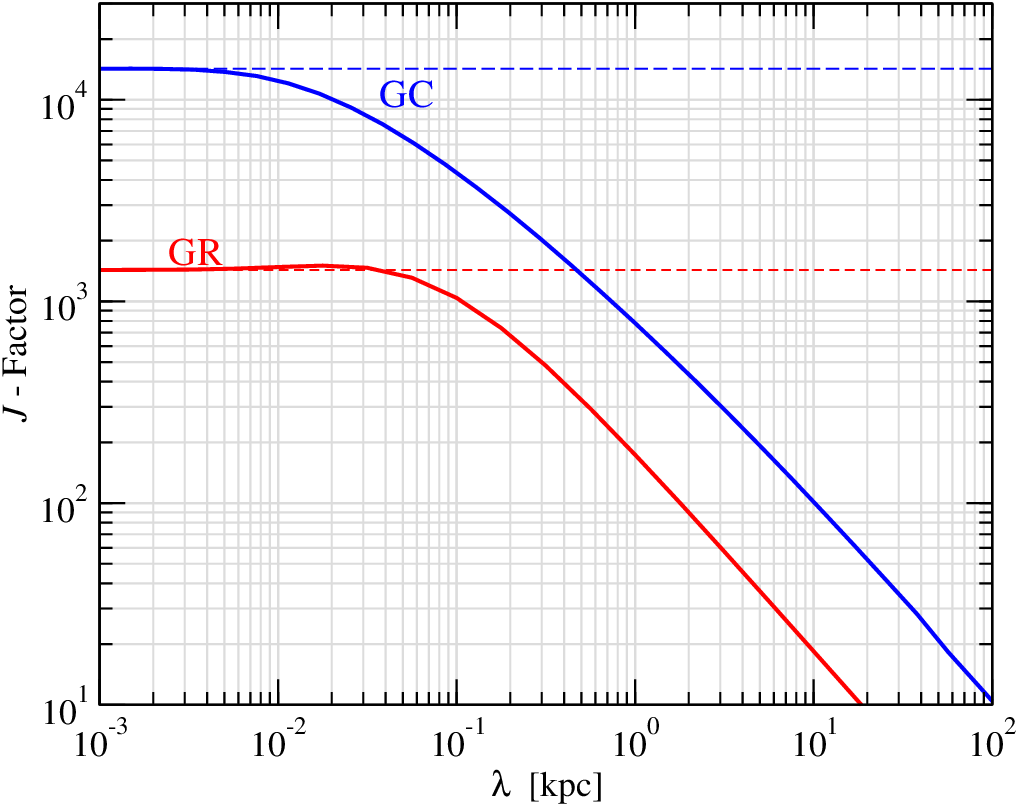}
\mycaption{\label{fig:J} $J$-factors for GC and GR (background subtracted as
  in \cite{Aharonian:2006au}) as a function of $\lambda$. Dashed lines
  correspond to the $J$-factors for $\lambda=0$.}
\end{figure}

\begin{table}
\centering 
\begin{tabular}{c|c@{\qquad}c@{\qquad}c@{\qquad}c}
\hline\hline
$\lambda$ [kpc]& 0 & 1 & 10 & 100 \\
\hline
GC & 14200 & 781 & 101 &  10.4 \\
GR &  1430 & 174 & 18  &   1.9 \\
\hline\hline
\end{tabular}
  \mycaption{\label{tab:J} $J$-factors for GC and GR (background subtracted
  as in \cite{Aharonian:2006au}) for various values of $\lambda$.}
\end{table}

Let us now discuss the dependence of the $J$-factor, Eq.~\ref{eq:J}, on
$\lambda$.  In contrast to the usual annihilation case, $J$ encodes not only
the astrophysical DM profile, but it depends now also on the particle
physics parameter $\lambda$ via $\rho_\mathrm{eff}$. This dependence is
shown in Fig.~\ref{fig:J} for the GC and GR regions observed by HESS, see
Eqs.~\ref{eq:GC} and \ref{eq:GR}. The figure shows that $J$ gets reduced as
soon as $\lambda$ becomes larger than the size $x$ of the observed region,
since then most of the $\phi$ decay outside of the observed region. For the
GC we have $x \simeq r_\odot \sqrt{\Delta\Omega_\mathrm{GC}/\pi} \approx
15$~pc, whereas for the GR we have $x\simeq r_\odot \sqrt{\Delta l \Delta b}
\approx 73$~pc, in agreement with the values of $\lambda$ where the curves
in Fig.~\ref{fig:J} start to deviate from the $J$-factor for $\lambda = 0$.
In Tab.~\ref{tab:J} we give $J$ for the GC and GR for some values of
$\lambda$. In the case of GR the $J$ factors also take into account that
HESS subtracted the background by comparing to a region outside the center,
as mentioned after Eq.~\ref{eq:GR2}. The quoted $J$ factors are then a
difference of the average fluxes from the two regions. For nonzero $\lambda$
this can lead to a further suppression of a factor of few.

\section{Application to recent cosmic ray data}
\label{sec:data}

In this section we apply our LLP scenario to the recent data from cosmic ray
experiments and discuss various other constraints. We describe our fit to
the electron and positron data from PAMELA, ATIC, FERMI and HESS in
Sec.~\ref{sec:el}, and show that such a fit can be made consistent with the
HESS GC and GR observations in Sec.~\ref{sec:HESS}. The bounds from
neutrinos are discussed in Sec.~\ref{sec:neutrino}.

\subsection{The electron-positron signals}
\label{sec:el}

The electron flux from DM annihilations (which is equal to the
positron flux) at $r_\odot$ is calculated as
\be\label{eq:F_el}
\frac{d\Phi_e}{dE_e} = \frac{v_e}{4\pi b(E_e)} \, \sigv \,
\frac{\rho_\mathrm{eff}^2(r_\odot)}{m_\chi^2} 
\int_{E_e}^{m_\chi/2} dE' \frac{dN_e}{dE'} \, I(\lambda_D(E_e,E')) \,,
\ee
with $b(E_e) = E_e^2/({\rm GeV}\, \tau_E)$, $\tau_E = 10^{16}$~s, and
the electron velocity $v_e \approx c$.  $dN_e/dE$ is the injection
spectrum for electrons (equal to the one for positrons) per $\phi$
decay, which we calculate by assuming that $\phi$ decays into
muons. Then $dN_e/dE$ is calculated from the decay of the muons by
using {\tt pythia-6.4.19} \cite{Sjostrand:2006za} taking into account
final state radiation. We provide analytic parameterizations of the
injection spectra in Appendix~\ref{app}. The diffusion length
$\lambda_D$ is given by $\lambda_D^2(E,E') = 4 K_0 \tau_E(E^{\delta
  -1} - {E'}^{\delta -1})/(1-\delta)$ with $E, E'$ in GeV and
throughout this work we assume the so-called MED propagation model
from~\cite{Delahaye:2007fr}, where $K_0 = 0.0112\,\rm kpc^2/Myr$ and
$\delta = 0.70$. The halo function $I(\lambda_D)$ is obtained as a
series of Bessel- and Fourier transforms of
$\rho^2_\mathrm{eff}(r)/\rho^2_\mathrm{eff}(r_\odot)$ {\bf in~\cite{Delahaye:2007fr}.   We 
instead write down a partial differential equation for $I(\lambda_D)$ and solve for it
numerically, which speeds up the computation greatly. Details are relegated to appendix B.} 
Note that the flux in Eq.~\ref{eq:F_el} is a factor 2 larger than in
the case of $\chi$ annihilations directly into muons, since we obtain
2 $\phi$'s for each $\chi\chi$ annihilation, each of them giving a
$\mu^+\mu^-$ pair.

We consider the measurement of the positron fraction $\Phi_{e^+}/(\Phi_{e^+}
+ \Phi_{e^-})$ from PAMELA~\cite{Adriani:2008zr}, where we use only the 9
data points above 6.8~GeV where the effect of solar modulation is expected
to be small. Then we use data on the sum of electrons and positrons
$(\Phi_{e^+} + \Phi_{e^-})$ from ATIC~\cite{atic}, FERMI~\cite{Abdo:2009zk},
and HESS~\cite{Collaboration:2008aaa, Aharonian:2009ah} (the effect of older
data with much larger errors, e.g. PPB-BETS~\cite{Torii:2008xu}, is
expected to be small, therefore they are not included in the fits). Since
ATIC and FERMI data are not consistent with each other we do not combine
them but present results using only either of the two. For ATIC we use the
combined data from ATIC 1, 2, and 4, see last reference in \cite{atic}. The
two HESS measurements~\cite{Collaboration:2008aaa} and
\cite{Aharonian:2009ah} overlap in the intermediate energy range. Therefore
we use only the first 4 data points from \cite{Aharonian:2009ah}, while in
the overlap region we take only the measurements of
\cite{Collaboration:2008aaa}, which have smaller errors. The overall energy
scale of FERMI and the two HESS data sets is varied within the quoted
uncertainties. When we fit ``PAMELA only'' also the lowest three data points
from ATIC are included in order to constrain the normalization of the
background fluxes.
For the astrophysical electron and positron background fluxes we use
the parameterization from~\cite{Baltz:1998xv} for the fluxes from
Galprop~\cite{Moskalenko:1997gh}. Following~\cite{Cirelli:2008id},
these fluxes are multiplied by $C_{e^\pm} E^{\alpha_{e^\pm}}$, where
in the fit we allow free normalization constants $C_{e^\pm}$ and
assume $\alpha_{e^\pm} = 0 \pm 0.05$ ($1\sigma$), independently for
$e^-$ and $e^+$.

\begin{figure}
\centering 
\includegraphics[width=0.9\textwidth]{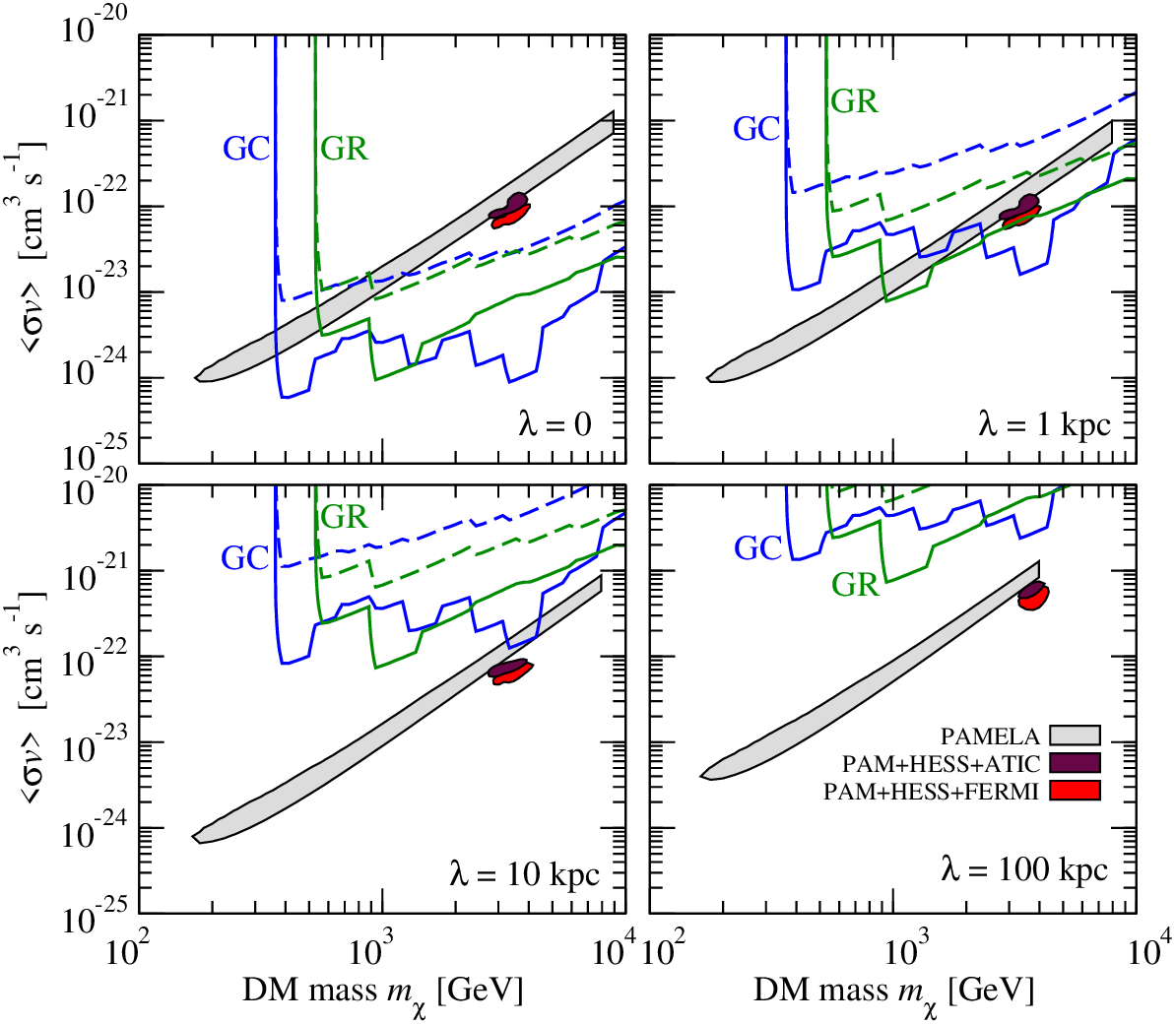} 
\mycaption{\label{fig:regions} Allowed regions at $3\sigma$ for PAMELA
  (gray), PAMELA+HESS+FERMI (red), and PAMELA+HESS+FERMI (dark red) and the
  constraints from HESS photon observations of the galactic center (GC) and
  galactic ridge (GR) for $\lambda =0, 1, 10, 100$~kpc. The solid (dashed)
  curves show HESS photon constraints at 90\%~CL with (at $3\sigma$ without)
  including a power law background in the fit, see text for details. The
  regions above the curves are excluded.}
\end{figure}

\begin{table}
\centering
\begin{tabular}{c|ccr|ccr|ccr}
\hline\hline
& \multicolumn{3}{|c}{PAMELA+HESS+ATIC}
& \multicolumn{3}{|c}{PAMELA+HESS+FERMI}
& \multicolumn{3}{|c}{PAMELA-only}\\
$\lambda$ [kpc]&
$m_\chi$ [TeV] & $\sigv \,[\rm cm^3s^{-1}]$ & $\chi^2_\mathrm{min}$ &
$m_\chi$ [TeV] & $\sigv \,[\rm cm^3s^{-1}]$ & $\chi^2_\mathrm{min}$ &
$m_\chi$ [TeV] & $\sigv \,[\rm cm^3s^{-1}]$ & $\chi^2_\mathrm{min}$ \\
\hline
   0 & 
   3.6 & $1.2\times 10^{-22}$ & 102.0  & 
   3.5 & $7.9\times 10^{-23}$ & 76.6   & 
   0.40& $3.2\times 10^{-24}$ & 7.3 \\   
   1 & 
   3.6 & $1.1\times 10^{-22}$ & 102.1 &  
   3.5 & $7.9\times 10^{-23}$ &  75.2 &  
   0.35& $2.5\times 10^{-24}$ &   7.1 \\ 
  10 & 
  3.5 & $7.9\times 10^{-23}$ & 112.9 & 
  3.6 & $6.9\times 10^{-23}$ &  82.9 & 
  0.40& $2.5\times 10^{-24}$ &   8.0\\ 
  100 & 
  3.7 & $6.3\times 10^{-22}$ & 89.1 & 
  3.7 & $5.0\times 10^{-22}$ & 86.2 & 
  0.32& $10\times 10^{-24}$  & 7.9 \\ 
\hline\hline
\end{tabular}
\mycaption{\label{tab:best-fit} Best fit values of $m_\chi$ and $\sigv$ and
   the corresponding $\chi^2_\mathrm{min}$ values for electron-positron data
   for some representative values of the $\phi$ decay length $\lambda$. The
   number of degrees of freedom are 38, 43, 10 for PAMELA+HESS+ATIC,
   PAMELA+HESS+FERMI, PAMELA-only, respectively.}
\end{table}

\begin{figure}
\centering 
\includegraphics[width=0.9\textwidth]{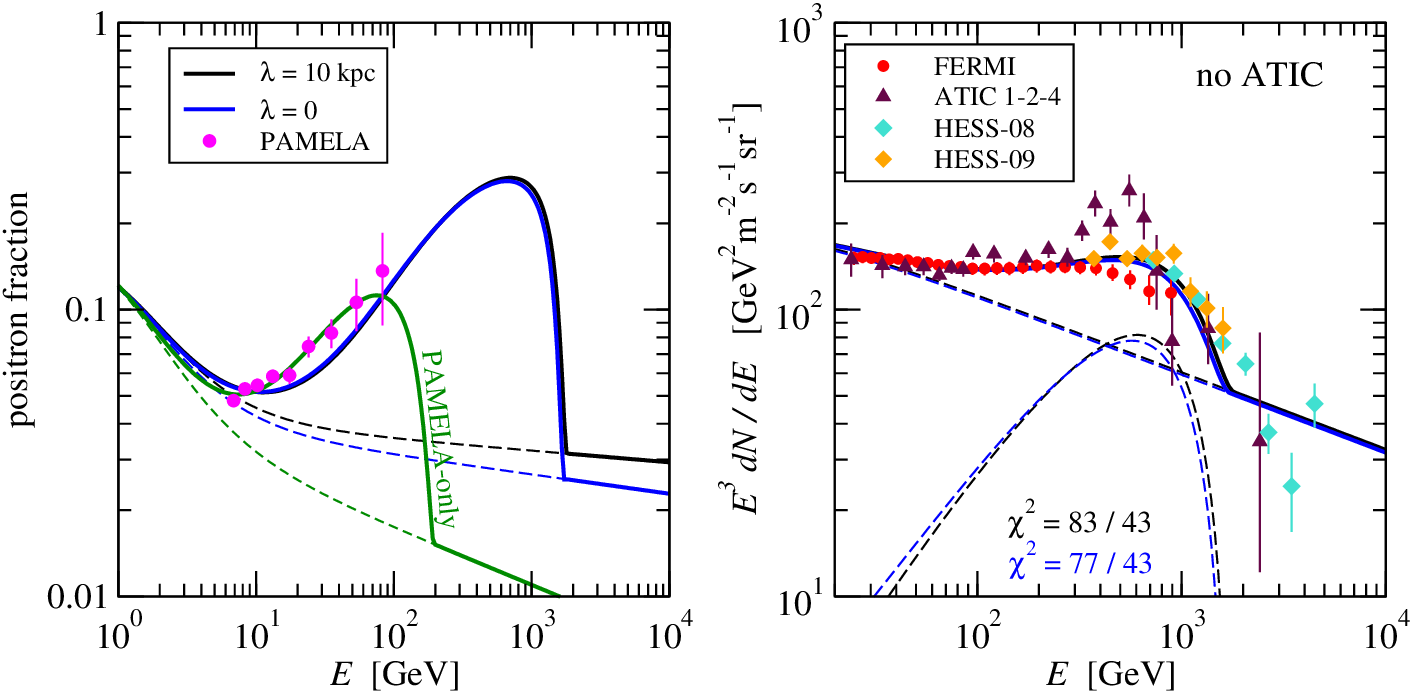} 
\includegraphics[width=0.9\textwidth]{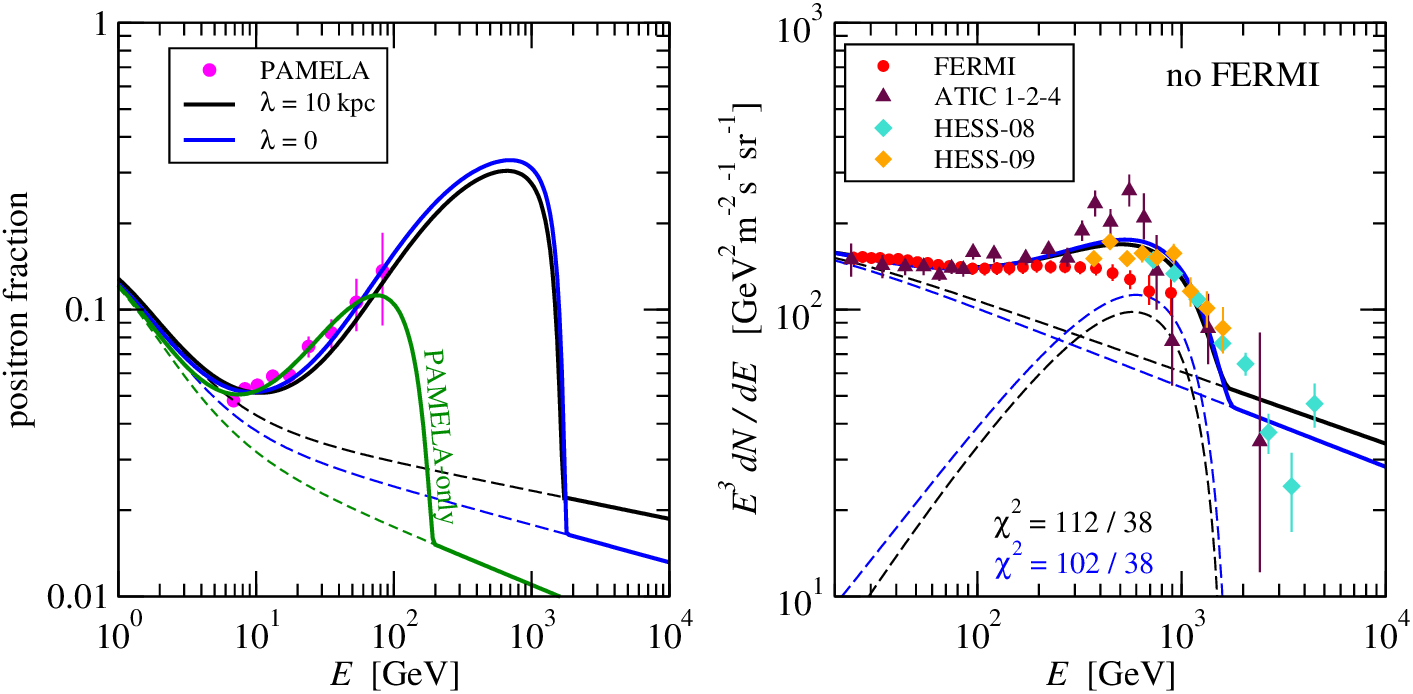} 
\mycaption{\label{fig:spectr} PAMELA data on the positron fraction (left)
  and the electron-positron data (right) compared to the predicted spectra
  for $\lambda=0$ (blue curves) and $\lambda=10$~kpc (black curves) at the
  best fit values given in Tab.~\ref{tab:best-fit}. Solid curves correspond
  to signal + background, whereas with the dashed curves we show background
  and signal (right panels only) components separately. Upper panels are for
  PAMELA+FERMI+HESS, lower panels for PAMELA+ATIC+HESS. The green curves in
  the left plots show the spectrum at the best fit to only PAMELA data for
  $\lambda = 10$~kpc.}
\end{figure}

The results of our fit to electron data are shown as the shaded regions in
Fig.~\ref{fig:regions} in the plane of the DM mass $m_\chi$ and the $\chi$
annihilation cross section $\sigv$ for four choices of the $\phi$ decay
length $\lambda$. These regions are defined by contours of $\Delta\chi^2 =
11.8$ with respect to the $\chi^2$ minimum ($3\sigma$ for 2~dof). Since
FERMI and ATIC data are inconsistent at about 3 $\sigma$ level, we perform
separate fits where one of the two data sets is excluded. The
corresponding best fit values are given in Tab.~\ref{tab:best-fit}, and the
fit to the data is shown in Fig.~\ref{fig:spectr}.

Irrespective of whether FERMI or ATIC data is used, the steep slope of the
high-energy HESS data provides a strong constraint on the DM mass. This then
leads to only a small change for the best fit values of $m_\chi$ and $\sigv$
in the two cases. If we fit only PAMELA data, however, there is a
degeneracy between $m_\chi$ and $\sigv$, and as long as the peak in the
electron spectrum is above the last PAMELA data point at 83~GeV a good fit
is obtained.  
As shown in Fig.~\ref{fig:regions} these results are basically
independent of the $\phi$ decay length $\lambda$, as long as this is
not much larger than the distance from us to the center of the
galaxy. The reason is that electrons and positrons are trapped in the
turbulent galactic magnetic field and the observed signal is dominated
by sources ``near by'', within several kpc. Therefore, they are not
very sensitive to the change of the source distribution from
$\rho^2(r)$ to $\rho^2_\mathrm{eff}(r)$, which is mostly important
close to the galactic center. Only if $\lambda$ becomes much larger
than $r_\odot \approx 8.5$~kpc the total electron production close to
us will be suppressed, which would require an increase in the
annihilation cross section to maintain the signal. This effect is
visible in Fig.~\ref{fig:regions} and Tab.~\ref{tab:best-fit} from the
results for $\lambda = 100$~kpc.

\subsection{Gamma ray constraints from HESS GC and GR observations}
\label{sec:HESS}

The photon fluxes predicted for given particle physics parameters
$m_\chi,\sigv,\lambda$ by Eq.~\ref{eq:F_gamma} can be compared with
observations. We use the gamma ray data from HESS observations of the
GC~\cite{Aharonian:2006wh} and the GR~\cite{Aharonian:2006au}.  These
data range from about 200~GeV to 20~TeV and are consistent with a
power law spectrum $\propto E_\gamma^{-\alpha}$ with $\alpha \approx
2.3$. In order to obtain bounds on DM parameters we adopt two
different strategies. Most conservative bounds can be obtained by
requiring that the signal predicted by DM must not exceed any data
point of the observed flux. We obtain these bounds (denoted by
``without background'') by excluding points in the parameter space
where the prediction exceeds $x + 3\sigma$ for any data point, where
$x$ is the observed flux and $\sigma$ its error bar. This leads to
very conservative bounds, since it requires that in the signal region
there is no astrophysical background, whereas in order to account for
the observed flux at energies where DM does not contribute some
astrophysical source has to be assumed. Therefore, we show also bounds
by using a second method, called ``with background''. Here we fit the
data with a power law background (with free normalization and power) +
the signal from DM.  For given $\lambda$ exclusion limits in the plane
of $m_\chi$ and $\sigv$ at 90\%~CL are obtained by the contours with
$\Delta\chi^2 =4.6$ with respect to the $\chi^2$ minimum.
The photon spectrum $dN_\gamma/dE_\gamma$ per $\phi$ decay used in
Eq.~\ref{eq:F_gamma} is calculated with {\tt pythia-6.4.19}
\cite{Sjostrand:2006za}, assuming that $\phi$ decays into
$\mu^+\mu^-+n\gamma$, see Appendix~\ref{app}.

The bounds from the HESS gamma ray observations are shown in
Fig.~\ref{fig:regions} together with the regions favored by
electron-positron data. Clearly, for $\lambda = 0$ the electron data are
inconsistent with the gamma ray constraints. The photon flux from the
galactic center gets reduced for finite $\lambda$, see Fig.~\ref{fig:J}, and
the bounds shift to larger values of $\sigv$ as $\lambda$ is increased.
Fig.~\ref{fig:regions} shows that for $\lambda \simeq 10$~kpc the regions
favored by electron-positron data are consistent with the HESS gamma ray
constraints, even for the 90\%~CL bounds with background. As one might
expect the bounds are also consistent with purely decaying dark
matter~\cite{Bertone:2007aw, Nardi:2008ix, Ibarra:2008jk} (and even in the
case where decaying dark matter is also allowed to annihilate
\cite{Cheung:2009si}). 
 
Apart from these high energy gamma ray observations additional information
can be obtained from less energetic photons. At energies around 10~GeV the
photon flux comes mainly from inverse Compton scattering on CMB photons,
star-light and dust-light, both for annihilating \cite{Chen:2009gz,
Borriello:2009fa, Lattanzi:2008qa, Harnik:2008uu} and decaying DM
\cite{Ishiwata:2009vx, Shirai:2009kh, Grajek:2008pg}. This leads to a
diffuse gamma ray signal, potentially observable by FERMI in the near
future. The recent diffuse gamma flux measurement by FERMI in the
$10^\circ - 20^\circ$ band above the galactic plane leads to a bound an
order of magnitude weaker than needed to probe the $m_\chi-\sigv$ region
preferred by electron/positron data even for the case of $\lambda=0$
\cite{Meade:2009iu}. Since the effective density is smaller in the inner
galactic region for $\lambda>0$, cf.\ Fig.~\ref{fig:rho_eff}, this bound is
expected to become even weaker for the LLP scenario. Looking away from the
inner galactic region the signal can, however, be stronger than for the
standard annihilating DM. To asses the impact of FERMI a more detailed
analysis is called for.

In addition to constraints from gamma rays, there are also strong
constraints due to synchrotron radiation of radio waves. These bounds
were extensively studied in~\cite{Bergstrom:2008ag, Bertone:2008xr,
  Zhang:2008tb, Ishiwata:2008qy} where it was found that annihilation
explanations of the positrons seem to be ruled out, unless the DM
profiles are made effectively less steep as in decays or as in LLP.
Thus it seems very plausible that for sufficiently large $\lambda$ LLP
can satisfy the bounds, though calculations along the lines
of~\cite{Bergstrom:2008ag,Bertone:2008xr} are in order. Constraints
using potential gamma ray signals from dwarf galaxies, on the other
hand, are less powerful \cite{Essig:2009jx,Pieri:2009zi}. Constraints
from galaxy clusters have been discussed in \cite{Yuan:2009yy}.

\subsection{Neutrinos from the galactic center}
\label{sec:neutrino}

Super-Kamiokande (SK) provides an upper limit on the upward going muon
flux from various extra-terrestrial sources~\cite{Desai:2004pq},
see~\cite{Abbasi:2009uz} for similar recent results from Ice Cube.
Since the LLP scenario predicts 8 neutrinos for each DM annihilation (from
the decay of the four muons from $\chi\chi\to\phi\phi\to 2\mu^+
2\mu^-$) the bound on upward going muons from the galactic center is
potentially relevant.  The neutrino induced muon flux can be
calculated as, see e.g.~\cite{Delaunay:2008pc}
\be\label{eq:mu-flux}
\Phi_\mu = \int_{E_\mathrm{thr}}^{m_\chi/2} dE_\nu 
\frac{d\Phi_\nu}{dE_\nu} 
\int_{E_\mathrm{thr}}^{E_\nu} dE_\mu \,
R_\mu(E_\mu)
\sum_{a=p,n} n_a \sum_{x=\nu,\bar\nu}
\frac{d\sigma_x^a(E_\nu)}{dE_\mu} \,.
\ee
Here,
\be
R_\mu(E_\mu) = \frac{1}{\rho \beta_\mu}
\ln\frac{\alpha_\mu + \beta_\mu E_\mu}{\alpha_\mu + \beta_\mu E_\mathrm{thr}}
\ee
is the range of a muon with energy $E_\mu$ until its energy drops
below $E_\mathrm{thr}$, for which we take the SK analysis threshold of
1.6~GeV, with $\alpha_\mu = 2\times 10^{-3}\,\rm GeV\,cm^2\,g^{-1}$,
$\beta_\mu = 4.2\times 10^{-6}\,\rm cm^2\,g^{-1}$, and $\rho$ is the
density of the material in $\rm g\,cm^{-3}$. Further, $n_a \approx r_a
\, \rho/m_p$ are the number densities of neutrons and protons with
$r_p \approx 5/9$, $r_n\approx 4/9$, and for the detection cross
section we use
\be
\frac{d\sigma_x^a(E_\nu)}{dE_\mu} \approx \frac{2m_p G_F^2}{\pi}
\left(A_x^a + B_x^a\frac{E_\mu^2}{E_\nu^2}\right)
\ee
with $A_\nu^{n,p} = 0.25, 0.15$, $B_\nu^{n,p} = 0.06, 0.04$, and
$A_{\bar\nu}^{n,p} = B_\nu^{p,n}$, $B_{\bar\nu}^{n,p} =
A_\nu^{p,n}$~\cite{Barger:2007xf}.

$d\Phi_\nu/dE_\nu$ is the flux of muon neutrinos arriving at the earth
within a solid angle $\Delta\Omega$. In our case the flux is equal for
neutrinos and anti-neutrinos, and it is given by
\be\label{eq:F_nu}
\frac{d\Phi_\nu}{dE_\nu} = 
\frac{\sigv}{4\pi} \,\frac{r_\odot \rho_\odot^2}{m_\chi^2}\,
\left(
  P_{\nu_e\to\nu_\mu}   \frac{dN_{\nu_e}}{dE_\nu} +
  P_{\nu_\mu\to\nu_\mu} \frac{dN_{\nu_\mu}}{dE_\nu} 
\right)
J \, \Delta\Omega \,.
\ee
The oscillation probabilities in terms of the lepton mixing matrix
elements $U_{\alpha i}$ are
$P_{\nu_e\to\nu_\mu} = \sum_{i=1}^3 |U_{ei}|^2|U_{\mu i}|^2 \approx
0.21$, $P_{\nu_\mu\to\nu_\mu} = \sum_{i=1}^3 |U_{\mu i}|^4 \approx
0.395$.  $dN_{\nu_e (\nu_\mu)}/dE_\nu$ is the spectrum of electron
(muon) neutrinos per $\phi$ decay (and integrates to 1). A
parameterization for it is given in Appendix~\ref{app}. The $J$-factor
in Eq.~\ref{eq:F_nu} is defined in the same way as for photons, see
Eq.~\ref{eq:J}, and therefore the neutrino flux will be reduced with
increasing the $\phi$ decay length $\lambda$ similar to the case of
photons.

\begin{figure}
\centering 
\includegraphics[width=0.6\textwidth]{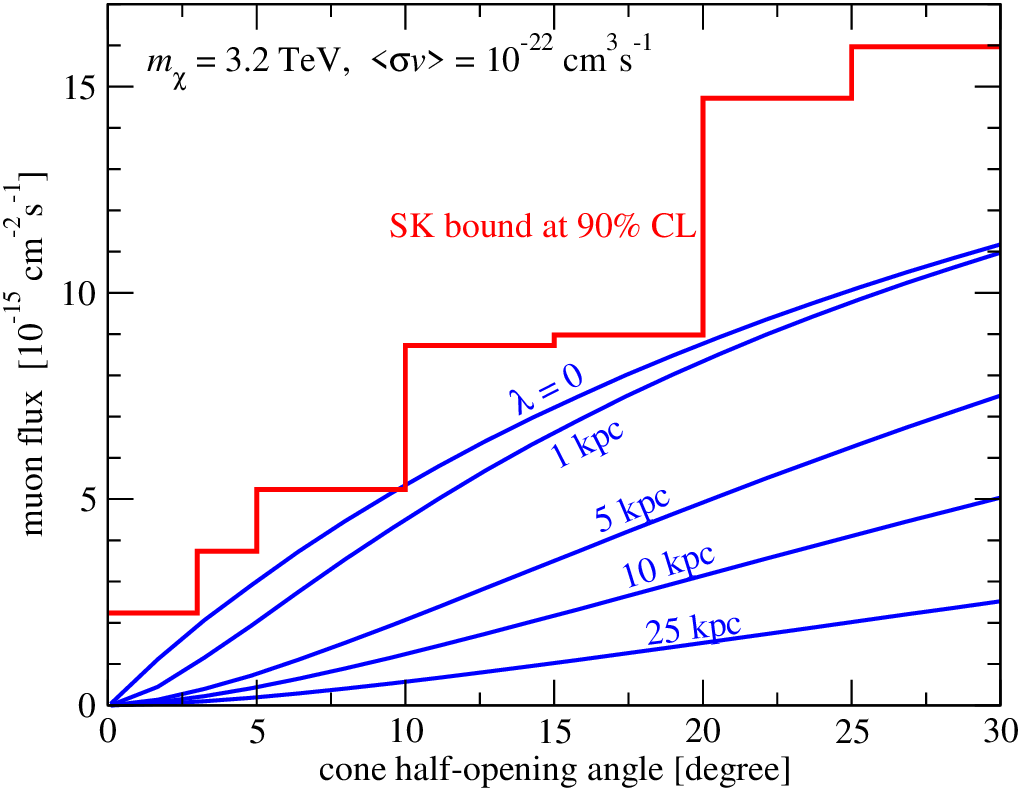} 
\mycaption{\label{fig:neutrino} Muon flux predicted for $m_\chi =
  3.2$~TeV, $\sigv = 10^{-22}\,\rm cm^3/s$ and various values of
  $\lambda$, compared to the 90\%~CL upper limit from
  Super-Kamiokande~\cite{Desai:2004pq}.}
\end{figure}

SK provides an upper bound on the muon flux from a cone around the
galactic center as a function of the cone half opening angle up to
$30^\circ$. In Fig.~\ref{fig:neutrino} we compare the SK upper bound
to the predicted muon fluxes for fixed $m_\chi$ and $\sigv$ as a
function of $\lambda$. We conclude that $\lambda = 0$ is close to the
present bound, while for $\lambda \gtrsim 5$~kpc the SK bound is
significantly relaxed. For SK neutrino bounds for decaying DM
see~\cite{Hisano:2008ah, Arvanitaki:2008hq, Liu:2008ci}.

Let us also mention that in the LLP scenario we do not expect any
observable neutrino flux from DM annihilations in the sun (or in the
earth), since $\phi$ is to a very good approximation stable at the
scale of the solar system. These neutrino fluxes will be exponentially
suppressed by the ratio of the sun--earth distance (or the earth
radius) to $\lambda$.

\section{Can $\chi$ be a thermal relic?}
\label{sec:relic}

Let us now address the issue of whether or not $\chi$ can be a thermal
relic. The LLP scenario, where $\chi$ annihilates first to
2 $\phi$'s and these then after some time decay to SM particles, faces
similar challenges  as direct   $\chi$'s annihilation into
SM particles.

First of all, for $\chi$ to be a thermal relic, the required $\chi$
annihilation cross section at the time of freeze-out in the early universe
is several orders of magnitudes too small to explain the PAMELA/ATIC/FERMI
cosmic ray anomaly. For $\chi$ with TeV mass, the cross section at the
freeze-out should be about $\langle \sigma v \rangle_{\rm FO} \sim 3\times
10^{-26}\,\rm cm^3 s^{-1}$, while as we have seen in section~\ref{sec:el},
the annihilation cross section that explains the cosmic ray anomaly, is
much larger, $\langle \sigma v \rangle_{\rm PA} \sim 10^{-22} \,\rm cm^3
s^{-1}$. One intriguing possibility which can explain the mismatch is that
there exists an attractive long range force between the dark matter
particles \cite{Hisano:2006nn}\footnote{Long range means here that the mass
of the force carrier is smaller than $M v$.}. This leads to an enhancement
of the cross section at small velocities by a factor of $1/v$. With a
typical velocity of $\chi$'s in the galactic halo $v/c\sim 10^{-3}$ this so
called ``Sommerfeld enhancement'' is roughly of the right size. In the LLP
scenario for the case of a non-relativistic $\phi$, there is a final phase
space suppression of the cross section for $\chi$'s annihilating in the
galactic halo compared to the early universe. However, larger Sommerfeld
enhancements are attainable, if there exists a bound state very close to
threshold \cite{MarchRussell:2008tu,Pospelov:2007mp}.

The Sommerfeld enhancement solution has several potential
phenomenological problems. For instance, if the Sommerfeld enhancement
worked to arbitrarily small velocities, then annihilation in
proto-halos could lead to a too large contribution to the diffuse
gamma ray background \cite{Kamionkowski:2008gj}. For massive enough
attractive force carriers this bound does not apply (such as GeV mass
force carriers of \cite{ArkaniHamed:2008qn}) as the enhancement saturates.
More importantly, the
highly energetic leptons and photons originating from $\chi$
annihilations at $T\lesssim 10$~keV in the early universe could lead to
photo-dissociation of light elements destroying the successful
predictions of standard big bang nucleosynthesis (BBN)
\cite{Hisano:2009rc}. The resulting bound on the annihilation cross
section for $\chi \chi \to \mu^+\mu^-$ is $\langle \sigma v
\rangle_{\rm FO} < 2.0\times 10^{-23}\,{\rm cm^3 s^{-1}}\times (m_\chi
/1 \, {\rm TeV})^{-1}$. This bound is already somewhat smaller than
needed for the explanation of the cosmic ray anomaly, see
Fig~\ref{fig:regions}. The bound itself was obtained assuming time
independent cross section, so a more detailed study in the framework
of models giving Sommerfeld enhancement (and in the LLP scenario) may
be warranted. However, since in our case $\phi$ is very long lived
compared to BBN time scales, one may expect that within the LLP
scenario there is no threat for BBN from rapid $\chi$ annihilations,
while there are constraints from the late decays of $\phi$'s, see
below.

There are other ways to ``boost'' the annihilation cross section in
the galactic halo without running into these phenomenological
problems. For instance the annihilation could go through a resonance
with mass of order $2m_\chi$ \cite{Ibe:2008ye,Guo:2009aj}.  This could
arise naturally in models of extra dimensions with linearly spaced KK
modes.  Another possibility would be a kination model
\cite{Salati:2002md, Rosati:2003yw}, where the expansion rate at the
time of decoupling is increased due to a rolling scalar field, leading
to a reduced relic density. Yet another possibility is that $\chi$ is
a product of a decay of a different meta-stable thermal relic
\cite{Fairbairn:2008fb}. Clearly, no boost factor is needed for
decaying DM, but the large decay time leads to interesting
model-building implications \cite{Arvanitaki:2008hq, Pospelov:2008rn,
  Chen:2008qs, Yin:2008bs}.

In the LLP scenario, where $\chi$ is a thermal relic and $\phi$ a
meta-stable thermal relic, there are other constraints coming from the
late decay of the $\phi$ particle. Phenomenologically, the LLP
scenario is interesting if $\phi$ travels a distance $\lambda\gtrsim
10$~kpc, because then it has a significant impact on the expected
photon flux from the galactic center as discussed in
section~\ref{sec:HESS}. This means that the $\phi$ life time is 
\begin{equation}\label{eq:tau}
\tau = \frac{\lambda}{c\beta\gamma} 
\simeq \frac{10^{12} \,{\rm s} }{\beta\gamma} 
\left(\frac{\lambda}{10\,{\rm kpc}}\right) \,. 
\end{equation}
Such late decaying relics can lead to
modifications of the light element abundances~\cite{Ellis:1990nb,
  Kawasaki:1994sc}. For TeV masses and lifetimes in this range, the
$\phi$ particle must have a relic density which is three orders of
magnitude smaller than the one for $\chi$, see figures~5 and 6 of
\cite{Kawasaki:1994sc}. For $\tau \gtrsim 10^{13}$~s there is an even
stronger bound on the $\phi$ abundance due to constraints on the
diffuse gamma ray background~\cite{Ellis:1990nb, Kribs:1996ac} (see
figure~7 of \cite{Kribs:1996ac}, where the bounds were derived for the
case of radiative decays, while the bounds for leptonic decays are
expected to be slightly weakened). The ratio $\sigma^\mathrm{ann}_\chi
/ \sigma^\mathrm{ann}_\phi$ must therefore be of order $10^{-3}$ or
smaller.  Consequently, if the couplings to quarks are not excessively
suppressed, $\phi$ pair production should be observable at LHC.  We
leave it as an open exercise in model building to embed the above
hierarchy of cross sections in a concrete model.

\section{Discussion and conclusions}
\label{sec:conclusion}

In this paper we introduced a scenario for cosmic ray production due
to dark matter (DM) which is an alternative to annihilations or decays
\footnote{A scenario where the dark matter both anhinilates and decays
  was propsosed in \cite{Cheung:2009si}.}. The scenario softens the
angular distribution of the gamma ray spectrum by having the DM
annihilate into long lived particles (LLP), that subsequently
decay to SM particles. We have demonstrated within the context of the
PAMELA/ATIC/FERMI positron and electron cosmic ray anomaly, that while
standard DM annihilation scenarios seem to be in conflict with the
gamma ray data due to large required cross sections $(\langle \sigma
v\rangle \sim 10^{-23} \rm \, cm^3 s^{-1}$), the prolongation of the
final state lifetimes in the LLP scenario can accommodate the data. It
should be emphasized that the gamma ray bounds are sensitive to the
dark matter profiles.  For less cuspy profiles the bounds would be
weakened. Here we have only considered the NFW profile which grows as
$1/r$ for small $r$.  Recent simulations \cite{Gnedin:2004cx,
  Kazantzidis:2004ee} that include baryons seem to indicate an even
faster rise for small $r$ which would only strengthen the gamma ray
bounds.

An attractive feature of DM annihilations is the direct relation
between the annihilation cross section for indirect DM detection and
the production cross section at collider experiments. In the LLP
scenario this connection is lost to some extent, since the annihilation
cross section of $\chi$ is not directly related to the production
cross section of $\phi$. However, assuming that $\chi$ and $\phi$ are
thermal relics, bounds from light element abundances and diffuse gamma
rays on decaying relics imply that the annihilation cross section of
$\phi$ has to be some orders of magnitude larger than the one of
$\chi$ in order to suppress its relic abundance, see discussion in
section~\ref{sec:relic}. Under the assumption of a non-negligible
coupling to hadrons this implies good prospects for $\phi$ production
at LHC (if kinematically accessible).

\begin{figure}
\centering 
\includegraphics[width=0.7\textwidth]{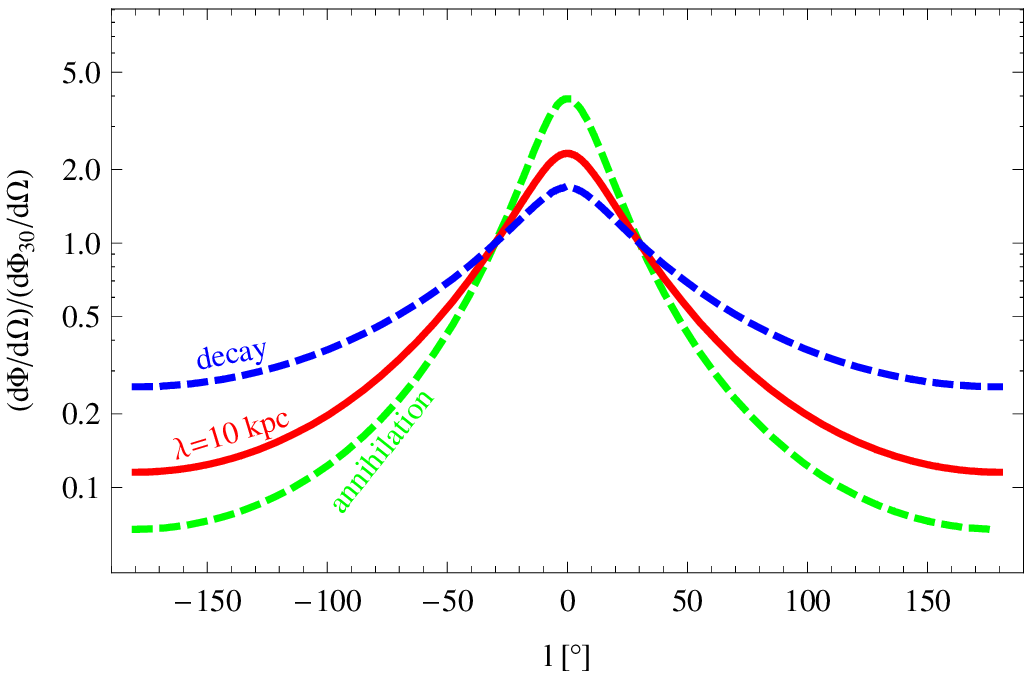} 
\mycaption{\label{angle} Angular distribution of the gamma ray flux
  for the LLP scenario with $\lambda=10$~kpc in comparison to the ones
  for DM decay and annihilation. We show the gamma ray flux for fixed
  galactic latitude $b=15^\circ$ as a function of the longitude
  $l$, normalized to its value at $30^\circ$.}
\end{figure}

It is interesting to ask how one would differentiate the LLP scenario
from DM decay. The FERMI satellite should be able to accurately
measure the angular distribution of the gamma ray flux.  This
distribution can be used to differentiate between decays and the LLP
scenario introduced in this paper.  Figure~\ref{angle} shows the gamma
ray flux at fixed galactic latitude $b=15$ degrees as a function of
the galactic longitude for the decay and the annihilation profiles
together with the one for the LLP scenario with $\lambda=10$~kpc. We
see that while the long lifetime reduces the slope compared to
annihilation it still cuts off faster than the decay profile.  An
analysis of the angular distribution for decaying dark matter was
performed in \cite{Bertone:2007aw} and it would be interesting to
extend this analysis to our scenario.

Another distinguishing feature of LLP will be the smearing out of
clumps.  If the only boost factor were due to clumps then this would
be a striking signal for dark matter. Typically the over-densities lie
towards the edges of galaxies where they are not tidally disrupted.
The long lived particle scenario would smooth out this striking
signal.

Here we have explored only the case of non-relativistic intermediate
states, which implies a near degeneracy between the DM and the
LLP. However, the softening of the angular distribution demonstrated
here, which allows one to avoid constraints from gamma ray bounds,
should be active in the case of relativistic annihilation products as
well. An investigation of this case is warranted.

Finally, let us speculate on the possible origin of the energy scale
corresponding to the typical life times of the LLP in our scenario,
which is set by our distance to the galactic center of order 10~kpc,
see Eq.~\ref{eq:tau}. If we assume that the decay of the LLP $\phi$
is governed by a dimension six operator suppressed by a high scale
$\Lambda$ one has roughly $\tau \simeq 16 \pi \Lambda^4/
m_\phi^5$. Then we find from Eq.~\ref{eq:tau} for the scale $\Lambda$
\begin{equation}
\Lambda \sim \frac{2\times 10^{12} \, {\rm GeV}}{(\beta\gamma)^{1/4}}
\left(\frac{\lambda}{10\,{\rm kpc}}\right)^{1/4} 
\left(\frac{m_\phi}{1\,{\rm TeV}}\right)^{5/4} \,. 
\end{equation}
We observe that the scale $10^{13}$~GeV corresponds roughly to the
seesaw scale for neutrino masses, since $v^2/\Lambda \sim 1$~eV for
$v\sim 100$~GeV and $\Lambda \sim 10^{13}$~GeV.\footnote{We mention
  that a similar argument for a dimension five operator would point to
  a scale $\Lambda$ above the Planck scale.}  This is particularly
intriguing in light of the leptonic nature of the currently observed
cosmic ray features and might indicate some common high scale physics
responsible for light neutrino masses and the LLP decay in our
scenario.

\subsection*{Acknowledgement}

We thank Christopher van Eldik for useful discussion on the HESS GR
analysis. T.S.\ acknowledges support from the Transregio
Sonderforschungsbereich TR27 ``Neutrinos and Beyond'' der Deutschen
Forschungsgemeinschaft.

\appendix 
\section{INJECTION SPECTRA}
\label{app}

Here we provide injection spectra for photons, electrons, and
neutrinos resulting from the decay $\phi \to \mu^+\mu^-$ in $\phi$'s
rest frame.\footnote{The spectra apply also for the case of standard
  annihilations $\chi\chi\to\mu^+\mu^-$ by replacing $m_\phi \to
  2m_\chi$.}  Since we work in the limit where $\phi$'s are
non-relativistic, the provided injection spectra can be directly used
in the numerical analysis in section~\ref{sec:data}. The spectra are
obtained by fitting analytic functions to a Monte Carlo sample
generated with {\tt pythia-6.4.19}~\cite{Sjostrand:2006za}. In the
following we define $x = 2 E / m_\phi$ and $0\le x \le 1$.

\bigskip 

{\bf Photon injection spectra.}
The photon spectrum for the decay (or similarly annihilation) of a
heavy particle with mass $m_\phi$ into charged particles is well described
by the Weizs\"acker-Williams form (see e.g. \cite{Bergstrom:2008ag,
  Bell:2008vx})   
\be\label{WW} 
\frac{d N_\gamma}{dE_\gamma} = \frac{\alpha_{\rm em}}{ E_\gamma \pi} [(1-x)^2+1] 
\ln\left[\frac{m_\phi^2(1-x)}{m_l^2}\right] \,.
\ee 
To be specific we have assumed here decay into a lepton pair
$l$. Though in our work we consider only the muon final state we list
here for completeness also the photon spectrum for electron and tau
final states.  The results from Pythia for muon and electron final
states are very close to Eq.~\ref{WW}. In the range for $E_\gamma$
from 10~GeV to 500~GeV for 1~TeV decaying DM the Pythia spectrum is
0.8 to 0.6 of the Weizs\"acker-Williams form. We parameterize the
spectrum by
\be\label{WWPythia} 
\frac{d N_\gamma}{dE_\gamma} = A \frac{\alpha_{\rm em}}{E_\gamma \pi}
\left(\ln\left[\frac{m_\phi^2(1-x)}{m_l^2}\right]+B\right)[(1-x)^2+C] \,.
\ee 
The coefficients in the above formula depend on the value of
$m_\phi$. We parameterize this dependence by 
\be\label{Kdependence}
K(m_\phi) = K_0 \left(\frac{m_\phi}{1{\rm TeV}}\right)^{\delta_K} \,, 
\ee
where $K=A,B,C$. Fitting Eq.~\ref{WWPythia} to the spectrum generated
with Pythia we find for $m_\phi\in[10^2,10^4] \, {\rm GeV}$:
\begin{align}
e^+e^- \,\text{final state:} & \left\{ 
\begin{array}{r@{\,=\,}l}
(A_0,B_0,C_0) & (1.49197,-11.0576,0.688768) \\
(\delta_A,\delta_B,\delta_C) & (0.050141,0.235939,-0.00623121)
\end{array} \right. \,,\nonumber\\
\mu^+\mu^- \,\text{final state:} & \left\{
\begin{array}{r@{\,=\,}l}
(A_0,B_0,C_0) & (1.36693,-2.94774,0.517128) \\
(\delta_A,\delta_B,\delta_C) & (-0.0365572,0.412598,-0.0137187)
\end{array} \right. \,.
\end{align}
The photon spectrum is then described to better than $13\%$ ($18\%$
for $\mu^+\mu^-$) everywhere in the fitted range (mostly to better
than $5\%$). This was obtained from 6 values of $m_\phi$, for each of
the values with $10^7$ simulated decay events.

For the decay into $\tau^+\tau^-$ we have produced $10^7$ events in
Pythia. Since the taus are heavy and can decay, producing hadrons and
leptons, and these further decay or bremsstrahl photons, the 
Weizs\"acker-Williams form Eq.~\ref{WW} is no longer a good description of the
final photon spectrum. A form that gives a good description is
\be
\frac{d N_\gamma}{dE_\gamma} =
A_0 \frac{\alpha_{\rm em}}{ E_\gamma \pi}\exp[-(A_1 x+A_2 x^2)] +
\frac{B_0}{1\,{\rm GeV}}+\frac{B_1}{1\,{\rm GeV}} x \,, 
\ee 
with $A_i,B_i$ depending as in Eq.~\ref{Kdependence} on $m_\phi$ with
\be
\begin{array}{r@{\,=\,}l}
(A_0^0, A_1^0, A_2^0, B_1^0, B_2^0) & 
(283.066, -0.981786, 5.80398,-0.000155047,0.000145457)\\ 
(\delta_{A_0}, \delta_{A_1}, \delta_{A_2}, \delta_{B_1}, \delta_{B_2})
& 
(0.00118045,-0.0141779,-0.00573809,-0.966239,-0.964131) 
\end{array}
\ee
The error on the predicted spectra is below $15\%$ (in the first bin
it is $30\%$).

\bigskip 
{\bf Electron injection spectra.}
Consider a two-body decay $\phi \to \mu^+\mu^-$ and neglect the photon
emission. Then the electron spectrum can be obtained by boosting the
isotropic spectrum from the muon rest frame into the rest frame of
$\phi$. In our work we do include final state radiation and
therefore, we prefer to use the electron spectrum generated with
Pythia. We use the function
\be
\frac{d N_e}{d(E_e/1\,\rm GeV)} = A_0 \frac{\alpha_{\rm em}}{ \pi}
\exp[-(A_1 x+A_2 x^2)] +B_0+B_1 x \,,
\ee
with $A_i,B_i$ depending as in Eq.~\ref{Kdependence} on
$m_\phi$. Fitting this form to the generated spectrum we find
\be
\begin{array}{r@{\,=\,}l}
(A_0^0, A_1^0, A_2^0, B_0^0, B_1^0) & 
(-0.296635, 2.65121, 14.8445, 0.0042505, -0.00427157) \\
(\delta_{A_0}, \delta_{A_1}, \delta_{A_2}, \delta_{B_0},\delta_{B_1})
& (-1.01424, 0.017198, -0.0107585, -0.999819, -0.999819) 
\end{array}\,.
\ee
The error on the predicted spectra is below $7\%$.

\bigskip
{\bf Neutrino injection spectra.}
In each decay $\phi \to \mu^+\mu^-$ we have two neutrinos---$\nu_\mu$
and $\nu_e$---and two antineutrinos with the same spectra. It turns
out to be convenient to fit for the full neutrinos spectrum
($\nu_e+\nu_\mu$) and for the electron neutrino spectrum separately.
The muon neutrino spectrum is then the difference between the total
neutrino spectrum and the electron neutrino spectrum.  For the total
$\nu_e+\nu_\mu$ neutrino spectrum (normalized to 2) we use the
parameterization
\be
\frac{d N_\nu}{d(E_\nu/1\,\rm GeV)} = A_0 \frac{\alpha_{\rm em}}{ \pi}
\exp\left[-(A_1 x+A_2 x^2)\right] + B \,,
\ee
with $A_i,B$ depending as in Eq.~\ref{Kdependence} on $m_\phi$ and
\be
\begin{array}{r@{\,=\,}l}
(A_0^0, A_1^0, A_2^0, B^0) & (3.73892,0.0128514,2.0015,-0.00117002) \\ 
(\delta_{A_0}, \delta_{A_1}, \delta_{A_2}, \delta_B) & 
(-1.0019,-0.283851,0.00350415,-1.01054)
\end{array}
\ee
with the error on the predicted spectra below $4\%$ (but mostly below
$2\%$). And for the electron neutrino spectrum we use
\be
\frac{d N_\nu}{d(E_\nu/1 \, \rm GeV)} = A_0 \frac{\alpha_{\rm em}}{
  \pi}
\exp\left[-(A_1 x+A_2 x^2+A_{1/2}\sqrt{1-x})\right] + B,
\ee
with again $A_i,B$ depending as in Eq.~\ref{Kdependence} on $m_\phi$, and 
\be
\begin{array}{r@{\,=\,}l}
(A_0^0, A_1^0, A_2^0, B^0,A_{1/2}^0) &
(14.4735,0.90821,2.90886,-0.000653652,1.95634) \\
(\delta_{A_0},\delta_{A_1},\delta_{A_2},\delta_B,\delta_{A_{1/2}}) &
(-0.994217,-0.00194938,0.00048921,-0.985772,-0.000395335) 
\end{array} 
\ee
The $\nu_e$ spectrum is described to better than 5\%.

\section{Differential equation for $I(\lambda_D)$}
The propagation of positrons in the galaxy is described by a transport equation (see, e.g. \cite{Delahaye:2007fr})
\be\label{diffusion:eq}
\partial_t\Psi -\nabla\big(K(x,E)\nabla \Psi)-\partial_E \big(b(E)\Psi\big)=q(x,E),
\ee
with $\Psi(\vec x,E)$ the positron number density per energy interval and $q(\vec x,E)$ the positron source term. The loss coefficient due to Compton scattering is 
$b(E)=E_0 \epsilon^2/\tau_E$ where $\epsilon=E/E_0$ with $E_0=1$ GeV and $\tau_E=10^{16}$~s, while the transport through the turbulent galactic magnetic field is
approximated by a spatially constant diffusion coefficient $K(\vec x,E)=K_0 \epsilon^\delta$. For annihilating dark matter the positron injection source term is
\be
q(\vec x, E)=\kappa \Big(\frac{\rho(\vec x)}{\rho_0}\Big)^2 f(\epsilon),
\ee
while for LLP scenario $\rho_{\rm eff}$ instead of $\rho$ should be used (for decaying dark matter, on the other hand, the dependence on $\rho$ is linear).
The positrons flux at radial distance from galactic center $r$ and distance $z$ from the galactic plane is then given by
\be\label{ansatz}
\Psi(r,z,\epsilon)=\kappa \frac{\tau_E}{\epsilon^2}\int_\epsilon^\infty d\epsilon_S f(\epsilon_S) \tilde I(\lambda_D, r, z),
\ee
where the diffusion length is $\lambda_D=\sqrt{4 K_0 \tilde \tau}$, with $\tilde \tau=\tilde t-\tilde t_S$ and $\tilde t=\tau_E(\epsilon^{\delta-1}/(1-\delta))$. The function $\tilde I(\lambda_D, r, z)$ depends only on astrophysics, i.e. only on the DM halo profile. The function $I(\lambda_D)$ in Eq. \ref{eq:F_el} is then equal to 
$\tilde I(\lambda_D, r, z)$ at the position of the solar system
\be
I(\lambda_D)=\tilde I(\lambda_D, r_\odot, 0).
\ee
In \cite{Delahaye:2007fr} the function $\tilde I(\lambda_D, r, z)$ is given in terms of an infinite sum over Bessel and cosine functions, however this expansion converges very slowly. 

Alternatively, using the ansatz Eq. \ref{ansatz} in Eq. \ref{diffusion:eq}, one can write down a partial differential equation for $\tilde I$
\be
\nabla^2 \tilde I(\lambda_D, r, z) - \frac{2}{\lambda_D} \partial_{\lambda_D} \tilde I(\lambda_D, r, z)=0.
\ee
The boundary conditions 
\be
\begin{split}\label{boundary:cond}
\tilde I(0, r, z)&=\Big(\frac{\rho(r,z)}{\rho_0}\Big)^2,\\
\tilde I(\lambda_D, R_{\rm gal}, z)&=0,\\
\tilde I(\lambda_D, r, \pm L)&=0,
\end{split}
\ee
follow from a requirement that $\Psi(r,z,\epsilon)$ vanishes on the boundaries of the galactic disk with radius $R_{\rm gal}$ and thickness $2L$ (the first boundary condition in Eq. \ref{boundary:cond} has to be multiplied by a smooth step function near the galactic disk boundaries to make it consistent with the next two boundary conditions). The above partial differential equation can be solved  numerically easily using standard methods, for instance the method of lines. The solution is obtained on a personal computer in less then a minute for the whole galaxy, compared to several hours for each point, if the expansion in Bessel and cosine functions is used. With appropriately modified
boundary conditions the above partial differential equation can be used for LLP scenario as well as for decaying dark matter. 


\end{document}